\pdfoutput=1

\documentclass[10pt,a4paper]{article} 
\usepackage{graphicx}
\usepackage{amsmath, amssymb, amsfonts, nicefrac, mathtools, amsbsy}
\usepackage{hyperref}
\usepackage[authoryear]{natbib}
\usepackage{rotfloat}
\usepackage[a4paper]{geometry}

\begin{document}
    \title{Optimal adaptive two-stage designs for single-arm trials with binary endpoint}
    \author{Kevin Kunzmann\thanks{University of Heidelberg, Institute for Medical Biometry and Informatics, INF 130.3, 69120 Heidelberg, Germany. \textit{Correspondence to}: kunzmann@imbi.uni-heidelberg.de}\ \ and Meinhard Kieser\footnotemark[1]}
    \date{\today} 
    
    \maketitle
    
\begin{abstract}
    Minimizing the number of patients exposed to potentially harmful drugs in
    early oncological trials is a major concern during planning. 
    Adaptive designs account for the inherent uncertainty about the true effect size
    by determining the final sample size within an ongoing trial after an interim 
    look at the data.
    We formulate the problem of finding adaptive designs which minimize expected
    sample size under the null hypothesis for single-arm trials with binary 
    outcome as an integer linear program.
    This representation can be used to identify optimal adaptive designs which 
    improve previous designs in two ways: 
    Firstly, designs can be found exhibiting lower expected sample size under the
    null hypothesis than those provided by previous algorithms.
    Secondly, we explain how integer programming techniques can be exploited to
    remove pathologies of the optimal and previous solutions arising from the 
    discrete nature of the underlying statistics.
    The resulting designs are both efficient in terms of expected sample size 
    under the null hypothesis and well interpretable. 
\end{abstract}

\maketitle

\section{Introduction}
\label{sec:intro}
Early phase II studies in clinical oncology are conducted after investigation of
safety and dosage of a new anti-tumor agent in preceding phase I studies. 
The objective of such early phase II trials is to identify substances that show 
promising anti-tumor activity warranting further research in larger phase II or
confirmatory phase III studies.
Usually, early phase II trials in clinical oncology are designed as single-arm 
studies with a binary endpoint indicating whether patients had substantial
tumor remission or, at least, no progression during a defined follow-up period.
Let $\rho$ be the probability of observing a favorable outcome (response) of the binary endpoint.
The main interest lies in testing the null hypothesis $\mathcal{H}_0:\rho 
\leq \rho_0$ for some pre-specified value $\rho_0$ chosen as the maximal value 
still considered clinically uninteresting.
Due to the early stage of clinical research at which this type of study is 
conducted, there is usually a high degree of
uncertainty about the magnitude of $\rho$.
In order to compensate for this uncertainty during planning, 
group sequential designs \citep{Jennison2000a} can be employed which allow 
early stopping for futility
or efficacy at pre-defined stages.
The large logistical effort of conducting interim analyses and the 
long follow-up times for oncological endpoints render multiple stages or 
strictly sequential testing impractical.
Instead, designs involving a single interim analysis after observing a 
pre-specified number of patients are viable compromises between the requirements
of clinical practice and the desire for the option of early stopping for either
futility or efficacy.
In his seminal paper, \cite{Simon1989} derived two types of two-stage 
group sequential designs which either minimize the expected sample size under 
the null hypothesis (Simon's optimal designs) or the maximal sample 
size (Simon's minimax designs).
Other optimality criteria have been proposed in the literature. 
For example, \cite{Shuster2002} aimed at optimizing the maximal expected sample size over the
complete range of success probabilities.
In the following, we focus on the optimality criterion of Simon's optimal 
designs.
This is due to its wide acceptance among practitioners and its intuitive
appeal: When testing potentially harmful substances \textit{in vivo}
one would like to minimize the number of patients exposed to non-active or
even harmful drugs. 
Should, on the other hand, the drug under investigation be beneficial, 
administering it to a larger number of patients within the study is unproblematic.
Therefore, minimizing expected sample size under the null hypothesis better
meets the ethical requirements than, say, minimizing expected sample size 
under a pre-specified alternative value $\rho_1> \rho_0$.
While two-stage designs allow early stopping in case of lower or higher 
treatment effect than assumed in the planning stage, they require that 
initially defined sample sizes and decision rules are to be followed strictly in order to assure control of the type one error rate.
During the last years, the development of more flexible methods for 
reacting to the observed interim outcome of a trial
have lead to an increasing research effort concerning so-called adaptive designs \citep{Bauer2015}.
This class of designs allows not only to stop early for futility or efficacy
but also to adjust the sample size of the second stage based on the observed interim outcome as long as the conditional error of 
the initially planned design is preserved.
Transferring the conditional error function principle \citep{Muller2004} to
discrete data, \cite{Englert2012a, Englert2012} derived flexible designs for
single-arm studies with binary endpoints.
Furthermore, formulating the decision rules in terms of the discrete conditional
error function results in counterparts of Simon's designs that are flexible and
at the same time as least as efficient if the pre-defined sample size is not
changed.
The same theoretical framework can be used to find sample size adaptation rules
that are optimal with respect to, e.g., the expected sample size under the null
hypothesis thus extending Simon's designs in a natural way. 
Previous approaches to the problem of finding optimal adaptive designs in the
sense of \cite{Simon1989} needed to impose technical constraints in order to
render the optimization feasible. 
\cite{Englert2013} imposed the additional constraint that the conditional error
function of the design should be non-decreasing in the number of observed 
positive outcomes after stage one (\mbox{EK designs}). 
While it might be intuitive to `shift' type one error to more promising stage-one outcomes, this was primarily a constraint for technical convenience as it 
also guarantees consistent regions for stopping for futility (conditional error of 0) or efficacy (conditional error of 1).
Yet, in some situations, the conditional error constraint is not sufficient to prevent the sample size function of the designs from being shaped oddly.
For example, the optimal design for $\rho_0=0.5$, a fixed alternative value of $0.7$, and $\alpha=0.05,\beta=0.1$ in \cite{Englert2013} requires 47 subjects for stage two upon observing 13 responses out of 20 subjects in stage one, 44 for 14 responses but 47 again for 15 responses (Table~1 in \cite{Englert2013}).
This non-smooth sample size function resulting from the discreteness of the underlying distribution is unintuitive and might impede adoption of these designs in practice.

The goal of this paper is two-fold. Firstly, we demonstrate that the problem can be solved in feasible time without any additional technical constraints using binary linear programming, cf. \cite{Garfinkel1972, Conforti2014}.
Subsequently, we investigate how restrictive the monotone conditional error constraint of Englert and Kieser is in practice and suggest a novel approach to 
obtaining `nice' solutions which prevent the potential pathologies of sample size function with negligible increase of the expected sample size under the null hypothesis.

The remainder of this paper is organized as follows. 
In Section~\ref{sec:notation}, we introduce the notation used throughout the
paper and describe the optimization problem for Simon's optimality criterion in
an adaptive two-stage setting.
Section~\ref{sec:formulation} explains how the problem can be formulated as a binary 
linear program with minimal constraints,
and Section~\ref{sec:tuning} explains the link to previous algorithms and explores the possibility of improving the quality 
of the solutions obtained with respect to various aspects. 
A numerical comparison for a range of commonly 
encountered  parameter configurations is given in Section \ref{sec:results}.
The discussion in Section \ref{sec:discussion} highlights the main 
differences of our approach to previous work and gives a prospect of possible 
extensions.

\section{Notation}
\label{sec:notation}
Throughout this paper, we consider two-stage single-arm clinical
trial designs with binary endpoint.
A pre-defined number of patients is enrolled during the first stage. Based on 
the number of observed responses and a pre-defined
sample size function, it is then decided how many patients are included in the 
second stage.
It is possible to stop early either for futility or for efficacy after the 
first stage.

Let $n_1$ be the number of patients enrolled in stage one, 
$X_1$ the number of responses observed in stage one, 
$X_2$ the number of responses observed in stage two and
$X=X_1 + X_2$ the overall observed number of such events.
The interest lies in testing $\mathcal{H}_0:\rho \leq \rho_0$ for a
pre-specified type one error rate $\alpha$ and a type two error rate $\beta$ at 
an alternative parameter value $\rho_1>\rho_0$.
Any two-stage design addressing this test problem can be described as 
a tuple $\big(n, c\big)$ of the total sample size function 
$n:\{0, 1, \ldots, n_1\} \to
\{n_1, n_1 + 1, \ldots, n_{max}\}$ and the
overall critical value function $c:\{0, 1, \ldots, n_1\} \to
\{0, 1, \ldots, n_{max}\}\cup\{-\infty, \infty\}$,
where $n_1$ is the number of patients pre-planned for the first stage and 
$n_{max}$ is the maximal total sample size. 
For any observed number of responses at interim $x_1$, 
the total sample size function $n(x_1)$ determines the number of patients 
enrolled in both stages and therefore the number needed for the second stage. 
After completing stage-two, the null hypothesis is rejected if and only if $X>c(x_1)$. 
Besides $n_{max}\geq n_1$, the functions $n(\cdot)$ and $c(\cdot)$ must fulfill 
the following consistency constraints for all $x_1 \in \{0, \ldots, n_1\}$
\begin{align}
    & c(x_1) \in \{-\infty, \infty\} \Leftrightarrow  n(x_1) = n_1 \label{eq:consistency}\\
    & n(x_1) > n_1 \Rightarrow n(x_1) > c(x_1) \geq x_1, \nonumber
\end{align}
which ensure that the final test decision is not yet fixed when continuing
to the second stage.

\section{Optimal two-stage design}
\label{sec:formulation}
Finding the optimal adaptive two-stage design for given type~one and type~two error rates $\alpha,\beta$ using Simon's classical optimality
criterion of minimal expected sample size under the null hypothesis can be
formulated as the following optimization problem
\begin{equation*}
    \begin{aligned}
        \underset{n_1,\,n(\cdot),\,c(\cdot)}{\text{minimize}}\quad & \boldsymbol{E}_{\rho_0}\big[\,n(X_1)\,\big] \\
        \text{subject to}\quad
        & \mathbb{P}_{\rho}\big[\,X > c(X_1)\,\big] \leq \alpha \quad  \forall\, \rho\leq\rho_0\\
        & \mathbb{P}_{\rho_1}\big[\,X > c(X_1)\,\big] \geq 1 - \beta.
    \end{aligned}
\end{equation*}
Note that all quantities involved are discrete and therefore specialized 
techniques are needed to solve the problem, 
which is not linear in the original variables $c(x_1),n(x_1), {x_1 = 0,\ldots,n_1}$. 
By reformulation as assignment problem using auxiliary variables it can, however, be stated as a binary linear program.
This class of problems can be solved efficiently with existing software.

Firstly, we simplify he problem by considering the optimization problem only for a particular value of $n_1$. 
The optimization over $n_1$ can then be performed by solving the conditional problem to optimality for every $n_1=1,2,\ldots$ until $n_1>\boldsymbol{E}_{\rho_0}\big[\,n(X_1)\,\big]$.
Let to this end for fixed $n_1$
\begin{align*}
    \big(y_{x_1,n_2,c}\big),
\end{align*}
$x_1=0\ldots n_1,n_2=0\ldots n_{max}-n_1, c=-\infty,0,\ldots,n_{max}-1,\infty$
be the three-dimensional binary assignment array with $y_{x_1,n_2,c}=1$ if and only if $n(x_1)-n_1=n_2$, and $c(x_1)=c$.
The fact that $n(\cdot)$ and $c(\cdot)$ must be valid functions of $x_1$ is easily represented by constraints of the form
\begin{align*}
    \sum_{n_2, c} y_{x_1,n_2,c} = 1 \quad \forall\, x_1 = 0\ldots n_1.
\end{align*}
The consistency constraints on $n(\cdot)$ and $c(\cdot)$ in equations (\ref{eq:consistency}) can be implemented as follows
\begin{align*}
    y_{x_1,n_2,c}=0&\quad \text{if}\ c(x_1)\in\{-\infty,\infty\}\wedge n_2 \in\{n_1 + 1,\ldots, n_{max}\} \\
    y_{x_1,n_1,c}=0&\quad \text{if}\ c \not\in\{-\infty,\infty\}\\
    y_{x_1,n_2,c}=0&\quad \text{if}\ n_2 > 0 \wedge \big(c<x_1\vee n_2 + n_1 \leq c\big).
\end{align*}
Additionally, let $\big(ce_{x_1,n_2,c}\big)$ be the corresponding three-dimensional array holding the respective conditional errors $ce(x_1)$, i.e., 
\begin{align}
    ce_{x_1,n_2,c} := &\ \mathbb{P}_{\rho_0}\big[X_1 + X_2 > c\,|\,X_1=x_1\big] \nonumber \\
    =&\ \mathbb{P}_{\rho_0}\big[X_2 > c - x_1\big],
\end{align}
and $\big(cp_{x_1,n_2,c}\big)$ the three-dimensional array holding the conditional power $cp(x_1)$ for each configuration, i.e.,
\begin{align}
    cp_{x_1,n_2,c} := &\ \mathbb{P}_{\rho_1}\big[X_1 + X_2 > c\,|\,X_1 = x_1\big] \nonumber \\
    = &\ \mathbb{P}_{\rho_1}\big[X_2>c-x_1\big].
\end{align}
Further, let $\big(n_{x_1,n_2,c}\big)$ be the three-dimensional array of corresponding sample sizes, i.e.,
\begin{align}
    n_{x_1,n_2,c} := n_2 + n_1.
\end{align}
The objective criterion can then be expressed as
\begin{align}
    \underset{\big(y_{x_1,n_2,c}\big)}{\text{minimize}} \quad \sum_{x_1,n_2,c} \mathbb{P}_{\rho_0}\big[X_1=x_1\big]\cdot n_{x_1,n_2,c}\cdot y_{x_1,n_2,c}
\end{align}
which is linear in the binary assignment variables $y_{x_1,n_2,c}$.
An overall power of $1-\beta$ at $\rho_1$ is guaranteed by the linear constraint
\begin{align}
    &\ \sum_{x_1,n_2,c} \mathbb{P}_{\rho_1}\big[X_1=x_1\big]\cdot cp_{x_1,n_2,c}\cdot y_{x_1,n_2,c} \geq 1-\beta.
\end{align}
Controlling the overall maximal type~one~error rate at $\alpha$ is more complicated.
Intuitively, the type~one~error rate should be largest at the boundary of the null hypothesis in which case the linear constraint
\begin{align}
    \sum_{x_1,n_2,c} \mathbb{P}_{\rho_0}\big[X_1=x_1\big]\cdot ce_{x_1,n_2,c}\cdot y_{x_1,n_2,c} \leq \alpha
\end{align}
would be sufficient to maintain type~one~error rate control.
Yet, we were unable to prove that this constraint is sufficient to achieve strict type~one~error rate control on $\mathcal{H}_0$.
We therefore resort to solving the problem only controlling the type~one~error rate at the boundary of the null hypothesis and numerically verify strict type~one~error rate control for the solutions obtained.
For all situations considered here the resulting designs maintained strict type~one~error rate control.

\begin{figure}
    \centerline{
        \begin{minipage}{\textwidth}
            \centering
            \includegraphics[width=\textwidth]{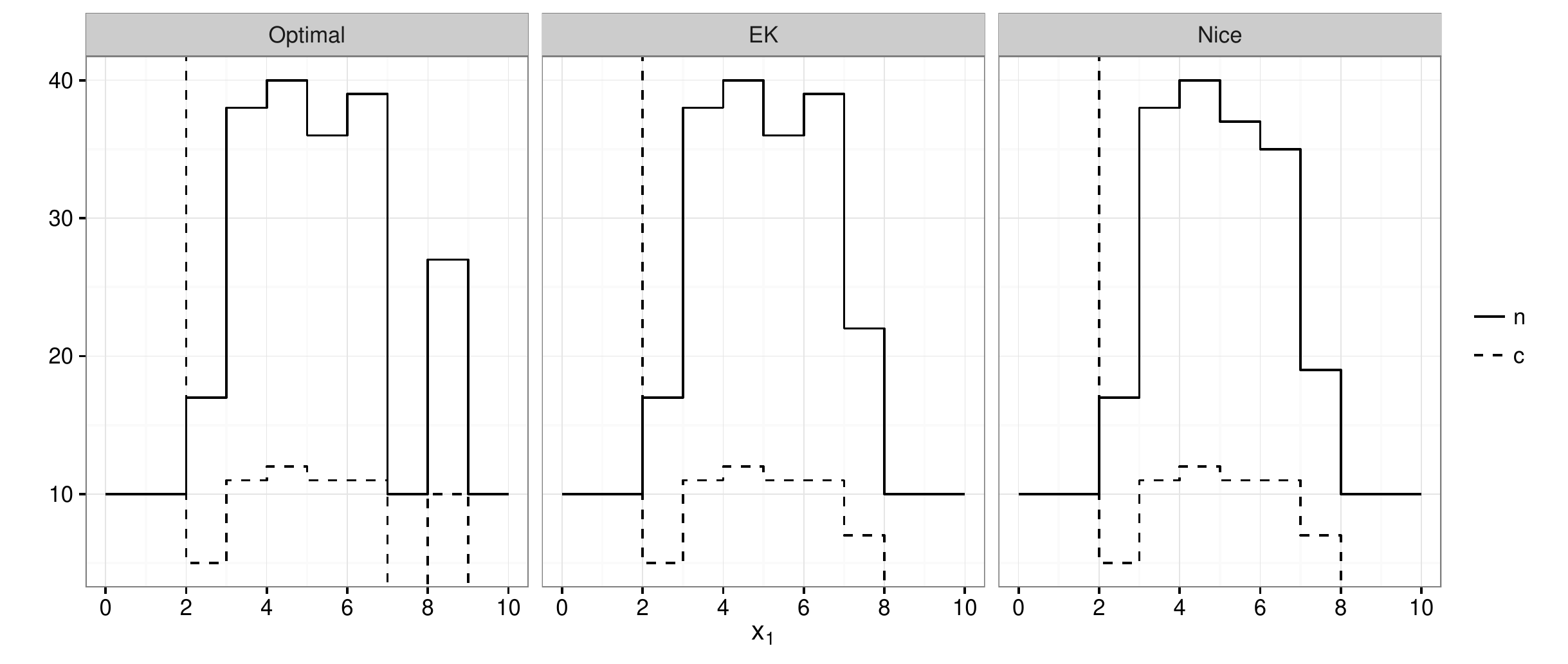}
            \includegraphics[width=\textwidth]{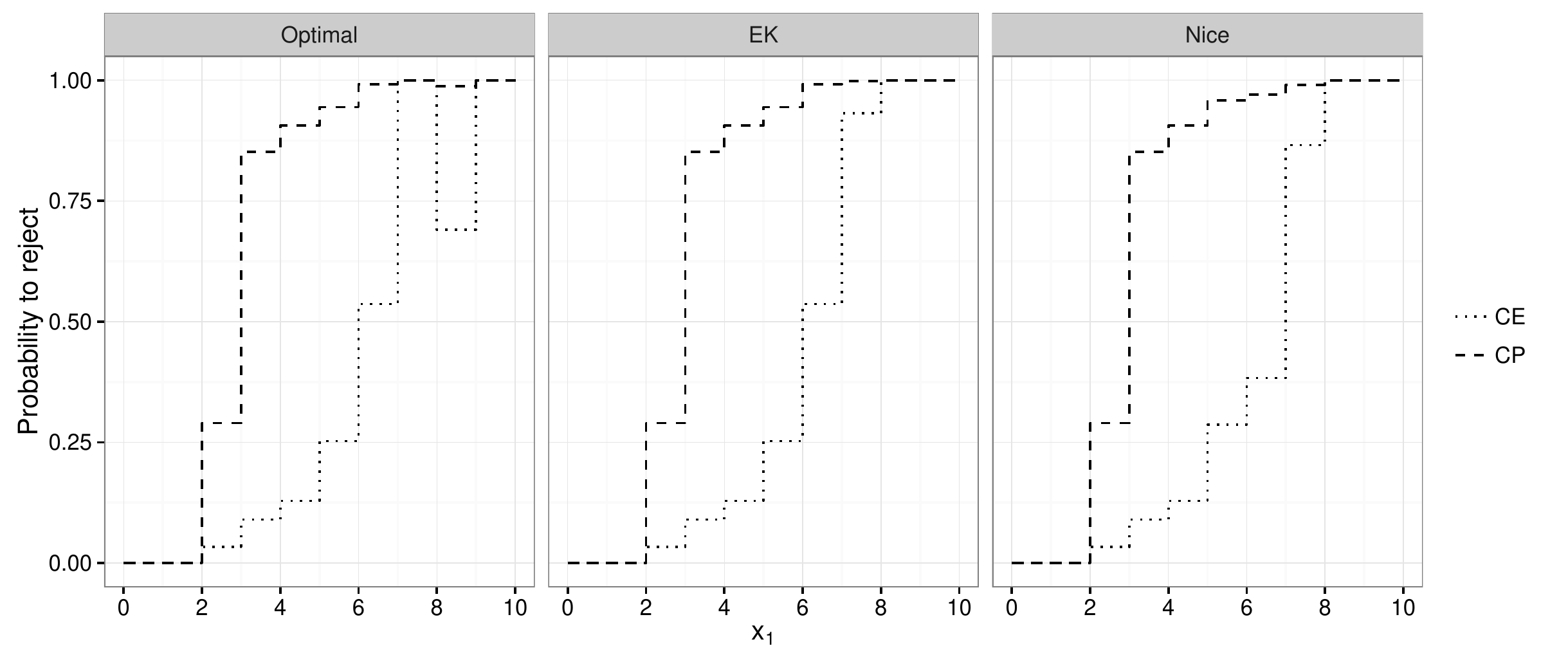}
        \end{minipage}
    }
    \caption{Sample size function $n(\cdot)$, critical value function $c(\cdot)$, conditional error function $ce(\cdot)$, and conditional power function $cp(\cdot)$ of the discussed adaptive designs for the example given in Section~\ref{sec:formulation}.}
    \label{fig:exampleDesigns}
\end{figure}

As an example, let $n_1=10, n_{max} = 40, \alpha=0.05, \beta=0.2, \rho_0=0.2$ and $\rho_1=0.4$.  
Table \ref{tab:examples} and Figure \ref{fig:exampleDesigns} show the
optimal adaptive design (`optimal') besides various modifications using additional constraints which are discussed in Section~\ref{sec:tuning}.
The optimal design has an expected sample size of 21.241. Although it violates monotonicity of $ce(\cdot)$ it still controls the type~one~error rate at $0.05$, which we checked numerically.
Therefore, the example shows that a monotone conditional error function is not a necessary condition for strict type~one~error rate control.
Also note that the stopping regions of the optimal design are not contiguous. 
In fact, the design stops for futility when observing 15 out of 15 responses in
stage one.
Although optimal in the sense of the specified criterion, such a behavior is not acceptable in practice and needs to be addressed.
In Section~\ref{sec:tuning} we first illustrate how the solutions of Englert~and~Kieser can be re-produced within the framework presented here and then explore alternative options for regularization of the
optimization problem.
\begin{table}
    \centering
    \caption{Sample size function $n(\cdot)$ and critical value function $c(\cdot)$ of the unconstrained (`optimal'), the monotone conditional error (`EK'), and the unimodal sample size function (`nice') adaptive designs for the example given in Section~\ref{sec:formulation}.}
    \label{tab:examples}
    \begin{tabular*}{\textwidth}{@{\extracolsep{\fill}}rrccccccccccc}
        \hline
        Design Type & $x_1$: & 0 & 1 & 2 & 3 & 4 & 5 & 6 & 7 & 8 & 9 & 10 \\
        \hline
        Optimal & $n(x_1)$: & 10 & 10 & 17 & 38 & 40 & 36 & 39 & 10 & 27 & 10 & 10 \\
        &$c(x_1)$: & $\infty$ & $\infty$ & 5 & 11 & 12 & 11 & 11 & -$\infty$ & 10 & - $\infty$ & - $\infty$ \\
        \rule{0pt}{4ex} 
        EK &$n(x_1)$: & 10 & 10 & 17 & 38 & 40 & 36 & 39 & 22 & 10 & 10 & 10 \\
        &$c(x_1)$: & $\infty$ & $\infty$ & 5 & 11 & 12 & 11 & 11 & 7 & -$\infty$ & -$\infty$ & -$\infty$ \\
        \rule{0pt}{4ex}
        Nice & $n(x_1)$: & 10 & 10 & 17 & 38 & 40 & 37 & 35 & 19 & 10 & 10 & 10 \\
        & $c(x_1)$: & $\infty$ & $\infty$ & 5 & 11 & 12 & 11 & 11 & 7 & -$\infty$ & -$\infty$ & -$\infty$ \\
        \hline
    \end{tabular*}
\end{table}

\section{Regularizing the solution with additional constraints} 
\label{sec:tuning}

\subsection{Monotone conditional error function} 

The solutions obtained by the branch-and-bound 
algorithm suggested by \cite{Englert2013} can be reproduced by adding constraints to the optimization problem that enforce a monotone conditional error function.
As a consequence, it is immediately clear that any solution obtained without
these constraints (`optimal' designs) will be at least as good as the one obtained by the EK 
algorithm in terms of the expected sample size under the null hypothesis.
Monotonicity of the conditional error function can be enforced by $n_1$ additional 
constraints of the form
\begin{align*}
    \sum_{n_2, c} ce_{x_1, n_2, c}\,\big(y_{x_1, n_2, c} -\ y_{x_1 - 1, n_2, c}\big)\geq 0
\end{align*}
for $x_1\in\{1,\ldots,n_1\}$.
As the example in Table~\ref{tab:examples} shows, however, the monotone 
conditional error function constraints on their own do not suffice to ensure 
`nice' solutions with sufficient practical appeal.
While the \mbox{EK design} achieves contiguous stopping regions, which are implied by a monotone conditional error function,
it does not guarantee a smooth sample size function.
The expected sample size under the null hypothesis is only slightly 
larger than the optimal solution's 
(21.250 vs. 20.241).

\subsection{Contiguous stopping regions} 

In order to obtain stopping regions which are connected to their respective boundary
(in case of stopping for futility to $x_1=0$ and in case of stopping for efficacy to $x_1=n_1$), one needs to enforce that for 
any fixed $x_1^*\in\{0,\ldots,n_1\}$ the following property holds:
$c(x_1^*) = \infty \Rightarrow c(x_1)=\infty$ for 
all $x_1 < x_1^*$ and \emph{vice versa} $c(x_1^*) = -\infty \Rightarrow c(x_1)=-\infty$ for 
all $x_1 > x_1^*$.
This set of conditional constraints can be formalized via $2\,n_1$ new 
binary variables $y^{fut}_{x_1}\in\{0,1\}$ for $x_1 \in\{1,\ldots,n_1\}$ and $y^{eff}_{x_1}\in\{0,1\}$ for $x_1\in\{0,\ldots,n_1 - 1\}$.
Let for  $x_1 \in\{1,\ldots,n_1\}$
\begin{align*}
    y_{x_1, n_1, \infty} - y^{fut}_{x_1} = 0.
\end{align*}
Then $y^{fut}_{x_1} = 1$ if and only if $c(x_1) = \infty$ (stopping for futility).
Adding a second constraint
\begin{align*}
    y_{x_1 - 1, n_1, \infty} - y^{fut}_{x_1} \geq 0
\end{align*}
enforces that $c(x_1 - 1)=\infty$ whenever $y^{fut}_{x_1} = 1$.
Therefore, by transitivity all stage-one outcomes smaller than $x_1$ must also 
lead to stopping for futility. 
Similarly, constraints for $y^{eff}_{x_1}$, $x_1 \in\{0,\ldots,n_1-1\}$ can
be constructed to ensure a contiguous stopping for efficacy region connected to $x_1=n_1$
\begin{align*}
    y_{x_1, n_1, -\infty} - y^{eff}_{x_1} &= 0\\
    y_{x_1 + 1, n_1, -\infty} - y^{eff}_{x_1} &\geq 0.
\end{align*}
While contiguous stopping regions are obviously implied by a monotonously increasing conditional
error function, the opposite is not true an therefore any design using only the contiguous stopping regions constraints instead of enforcing a monotone
$ce(\cdot)$ has more flexibility between the stopping regions.
In some cases, this additional flexibility might lead to a better performance in terms of minimal expected sample size under the null hypothesis.

\subsection{Unimodal sample size function} 

Motivated by the characteristics of the optimal solutions and the EK designs, we propose to resolve the issue of potentially unintuitive sample size functions by enforcing unimodality of the sample size function, which can be obtained by restricting the number of sign changes from 
positive to negative of the first order differences of $n(\cdot)$ to one.
This implies that at most one strict local maximum of the sample size function exists.
To this end, $n_1 + 1$ additional binary auxiliary variables $y_{x_1}^{\wedge}\in\{0,1\}, x_1\in\{0, \ldots, n_1\}$ and constraint sets 
\begin{align}
\sum_{n_2,c} n_{x_1',n_2,c}\cdot y_{x_1',n_2,c} - n_{(x_1 - 1),n_2,c}\cdot y_{(x_1-1),n_2,c} - 2\cdot n_{max}\cdot y_{i}^{\wedge}\geq -2\cdot n_{max},
\end{align}
$x_1'\in\{1,\ldots, x_1\}$ 
are needed. 
Whenever $y_{x_1}^{\wedge}=1$ these constraints enforce non-negative increments of $n_2(x_1')$ for all $x_1'<x_1$.
Similarly, non-positive increments for values larger than $x_1$ can be guaranteed by
\begin{align}
\sum_{n_2,c} n_{x_1',n_2,c}\cdot y_{x_1',n_2,c} - n_{(x_1'-1),n_2,c}\cdot y_{(x_1'-1),n_2,c} + 2\cdot n_{max}\cdot y_{x_1}^{\wedge}\leq 2\cdot n_{max}
\end{align}
for $x_1'\in\{x_1,\ldots,n_1\}$.
Finally, one must ensure that at least one of the above designed constraint sets is active in the solution. This can be achieved by 
\begin{align}
\sum_{x_1=0}^{n_1}y_{x_1}^{\wedge}\geq 1.
\end{align}
Jointly, the constraints for contiguous stopping regions and a unimodal sample size function result in designs which exhibit a `smooth' (unimodal) sample size function in all cases and guarantee that the stopping for efficacy and futility regions are contiguous and connected to their respective boundaries.
For the example given above the inflation of the expected sample 
size as compared to the optimal design is still negligible (21.252 vs. 21.241).

\section{Results}
\label{sec:results}

We compared the expected sample size under the respective null hypotheses for
four different sets of constraints over a range of parameter values for $\rho_0 = 0.1, 0.2, \ldots, 0.7$ and ${\rho_1=\rho_0 + 0.2}$. 
In all cases, we set $\alpha=0.05$ and $\beta=0.2$.
The search space for $n_1$ and $n_{max}$ was chosen with reference to Simon's 
original designs \citep{Simon1989}.
We allowed $n_{max}$ to be $10\%$ larger than the combined stage-one and stage-two sample size of the 
corresponding optimal design identified by Simon.
The search range for $n_1$ was chosen from $5$ up to $n_{max}-5$.
Table \ref{tab:results} shows the results. 
In all cases we numerically verified strict type~one~error rate control of the solutions.
Note that the small deviations from the figures reported in \cite{Englert2013}
originate from the fact that we did not need to restrict possible assignments by
specifying a minimal or maximal conditional power in order to render the 
optimization feasible.
All computations were conducted in the programming language Julia \citep{Bezanson2014} using its interface \citep{Lubin2015} to the commercial Gurobi solver \citep{Gurobi2015}. 
Graphics were produced using R \citep{r2015} and the ggplot2 package \citep{ggplot2009}.

The reference for our comparison is the optimal adaptive design without any
additional constraints. 
This design allows for maximal flexibility and must therefore exhibit the smallest expected sample size under the null hypothesis. 
Furthermore, we included designs equivalent to the ones obtained by Englert~and~Kieser by adding the monotone conditional error function constraint and `nice' designs which use the constraint-sets for contiguous stopping and
unimodal sample size function.
\begin{table}
    \centering
    \caption{
        Results for four different adaptive and Simon's designs using various combinations of the constraint sets discussed in Section~\ref{sec:tuning}.`\,$\cdots$' indicate the same result as to the left; $^*$: figures were taken from the original publication.
    }
    \label{tab:results}
    \begin{tabular*}{\textwidth}{@{\extracolsep{\fill}}cccccc}
        \hline
        \multicolumn{2}{c}{Parameters} & \multicolumn{4}{c}{$\boldsymbol{E}_{\rho_0}\big[\,n(X_1)\,\big]$}\\
        \cline{3-6}
        $\rho_0$ & $\rho_1$ & Optimal & EK & Nice & Simon's* \\
        \hline
        0.1 & 0.3 & 14.65107 & 14.72498 & $\cdots$ & 15.0 \\
        0.2 & 0.4 & 19.78640 & $\cdots$ & $\cdots$ & 20.6 \\
        0.3 & 0.5 & 23.02199 & $\cdots$ & 23.02448 & 23.6 \\
        0.4 & 0.6 & 24.08002 & 24.08640 & $\cdots$ & 24.5 \\
        0.5 & 0.7 & 22.94827 & $\cdots$ & 22.95923 & 23.5 \\
        0.6 & 0.8 & 19.71893 & $\cdots$ & $\cdots$ & 20.5 \\
        0.7 & 0.9 & 14.82367 & $\cdots$ & $\cdots$ & 14.8 \\
        \hline
    \end{tabular*}
\end{table}
The large search space with relatively high $n_{max}$ as compared to the example from Section~\ref{sec:formulation} results in the EK design mostly coinciding with the optimal and the `nice' design.
This indicates that the monotone conditional error function constraint is most restrictive when $n_{max}$ is relatively small, as it is the case in the example from Section~\ref{sec:formulation}.
In practice, however, $n_{max}$ need not always be chosen liberally due to operational constraints. 
It is therefore important that the unimodal sample size constraint guarantees  intuitive sample size functions in any situation.
For the parameter constellations considered here, both for $\rho_0=0.3$ and $\rho_0=0.5$, the EK and `nice' designs differ slightly despite the relatively large $n_{max}$ because the EK design's sample size function is not unimodal, cf. Table~\ref{tab:results}.

Overall, the differences in expected sample size between the optimal, EK, and nice designs are small for the situation considered here.
This indicates that the conditional error function approach of Englert and Kieser is not unnecessarily restrictive. 
However, the small differences to the nice designs demonstrate that the occasional issues with unintuitive sample size function can be resolved at minimal additional costs in terms of expected sample size under the null hypothesis.

\section{Discussion}
\label{sec:discussion}

We presented a framework for extending the classical optimality criterion of
\cite{Simon1989} to arbitrary adaptive two-stage designs improving previous work in two 
ways:
Firstly, we demonstrated how the problem can be formulated as a binary linear program which makes it amenable to solution by standardized and highly specialized software.
In this way, for the first time, we were able to solve the problem to optimality without any additional technical constraints.
Comparing the newly found optimal designs with the ones obtained previously by
\cite{Englert2013}, we conclude that the performance improvements in 
terms of expected sample size under the null hypothesis are almost negligible. 
Consequently, we conclude that the monotone conditional error constraint is not
unnecessarily restrictive.
Secondly, we utilized the binary linear program formulation to enforce certain desirable features of solutions obtained, which resolve problems arising due to the discreteness of the underlying statistics while preserving most of the advantages in terms of Simon's optimality criterion.
When comparing the optimal solutions without any additional constraints to the ones  obtained by adding `niceness' constraints such as unimodality of the
sample size function or contiguous stopping regions.
For many parameter settings the solutions coincide with the ones found by the algorithm of Englert and Kieser and where they differ due to a non-unimodal sample size function of the latter the performance loss in terms of expected sample size under the null hypothesis is extremely small.
Additionally, the fact that we formulated the problem in terms of a binary linear program allowed us to use commercial grade software for its solution. 
Thus, the solutions can be obtained very quickly which is a great improvement over naive implementations of the branch-and-bound algorithm.

Note that the binary linear programming framework presented in this paper also allows the addition of further constraints which can be used to tailor the solution's properties towards custom needs or preferences. 
For example, one might chose to require a conditional power of at least $1-\beta$ upon continuing to the second stage which would render sample size re-calculation based on conditional power unnecessary.

We considered expected sample size under the
null hypothesis as optimality criterion. 
However, we would like to mention that 
it is straightforward to
instead minimize the expected sample size under, e.g., some Bayesian prior 
distribution over $\rho$, 
which could then be seen as an extension of the ideas of \cite{Dong2012}.

\cite{Simon1989} also considered so-called `minimax designs' which minimize the maximal total sample size of the design.
Unfortunately, while it is theoretically possible to express minimax objective
functions in terms of a binary linear program \citep{bisshop2015}, 
this is not feasible in practice
as the number of additional constraints required is extremely large.
Although adaptive versions of Simon's minimax designs are thereby practically 
out of reach of the generic binary linear programming approach,
it is still possible to modify the objective function to favor designs with smaller maximal sample sizes. 
This can be achieved by minimizing $\boldsymbol{E}_{\rho_0}\big[\,n(X_1)^\gamma\,\big]$ for $\gamma>1$ or $\boldsymbol{E}_{\rho_0}\big[\operatorname{exp}\big(n(X_1)\big)\,\big]$ instead of $\boldsymbol{E}_{\rho_0}\big[\,n(X_1)\,\big]$.
Each of these objective functions is easily obtained by piecewise modifications
of the respective coefficients.
Alternatively, the parameter $n_{max}$ might be used more restrictively to
obtain a solution with acceptable maximal sample size.

Although all methods presented in this paper are developed for rate
comparisons of a single binomial random variable, an extension to more complex 
hypothesis tests based on, e.g., multinomial variables can be derived along the 
same lines.


\appendix


\subsection*{Acknowledgements}

The authors would like to thank the Deutsche Forschungsgemeinschaft 
(DFG) for supporting this research by grant \mbox{KI~708/1-3} and the 
contributors to the open source community for their invaluable efforts in 
creating the foundations for the software used for this paper.

\vspace*{-8pt}


%

\begin{thebibliography}{20}
    \bibitem[\protect\citeauthoryear{Bauer \textit{et~al.}}{2015}]{Bauer2015}
    Bauer, P., Bretz, F., Dragalin, V., K{\"{o}}nig, F. and Wassmer, G. (2015, published online). Twenty-Five Years of Confirmatory Adaptive Designs: Opportunities and Pitfalls. \textit{Statistics in Medicine}. DOI: 10.1002/sim.6472
    
    \bibitem[\protect\citeauthoryear{Bezanson \textit{et~al.}}{2014}]{Bezanson2014}
    Bezanson, J., Edelman, A., Karpinski, S. and Shah, V. B. (2014). Julia: A Fresh Approach to Numerical Computing. \textit{arxiv.org}. arXiv:1411.1607.
    
    \bibitem[\protect\citeauthoryear{Bisshop}{2015}]{bisshop2015}
    Bisshop, J. (2015). Linear Programming Tricks. In AIMMS - Optimization Modeling pp. 63-74. Lulu.com.
    
    \bibitem[\protect\citeauthoryear{Conforti, M., Cornu{\'{e}}jols, G., Zambelli, G.}{Conforti \emph{et~al.}}{2014}]{Conforti2014}
    Conforti, M., Cornu{\'{e}}jols, G., Zambelli, G. (2014). Integer Programming volume 271 of \textit{Graduate Texts in
        Mathematics}. Berlin: Springer.
    
    \bibitem[\protect\citeauthoryear{Dong \emph{et~al.}}{2012}]{Dong2012}
    Dong, G., Shih, W. J., Moore, D., Quan, H., Marcella, S. (2012). A Bayesian-Frequentist Two-Stage Single-Arm Phase II Clinical Trial Design. \textit{Statistics in Medicine} \textbf{31}, 2055--2067.
    
    \bibitem[\protect\citeauthoryear{Englert and Kieser}{2012\natexlab{a}}]{Englert2012}
    Englert, S., Kieser, M. (2012{\natexlab{a}}). Adaptive Designs for Single-Arm Phase II Trials in Oncology. \textit{Pharmaceutical Statistics} \textbf{11}, 241--249.
    
    \bibitem[\protect\citeauthoryear{Englert and Kieser}{2012{\natexlab{b}}}]{Englert2012a}
    Englert, S., Kieser, M. (2012{\natexlab{b}}). Improving the Flexibility and Efficiency of Phase II Designs for Oncology Trials. \textit{Biometrics} \textbf{68}, 886--892.
    
    \bibitem[\protect\citeauthoryear{Englert and Kieser}{2013}]{Englert2013}
    Englert, S., Kieser, M. (2013). Optimal Adaptive Two-Stage Designs for Phase II Cancer Clinical Trials. \textit{Biometrical Journal} \textbf{55}, 955--968.
    
    \bibitem[\protect\citeauthoryear{Garfinkel and Nemhauser}{1972}]{Garfinkel1972}
    Garfinkel, R., Nemhauser, G. (1972). Integer Programming. New York: Wiley.
    
    \bibitem[\protect\citeauthoryear{Gurobi Optimization, Inc.}{2015}]{Gurobi2015}
    {Gurobi Optimization, Inc.}. (2015). Gurobi Optimizer Reference Manual. http://www.gurobi.com
    
    \bibitem[\protect\citeauthoryear{Jennison and Turnbull}{1999}]{Jennison2000a}
    Jennison, C., Turnbull, B. W. (1999). Group Sequential Methods with Applications to Clinical Trials. 1st edn. Chapman and Hall/CRC.

    \bibitem[\protect\citeauthoryear{Lubin and Dunning}{2015}]{Lubin2015}
    Lubin, M. and Dunning, I. (2015). Computing in Operations Research Using Julia. \textit{INFORMS Journal on Computing} \textbf{27}, pp. 238--248.
    
    \bibitem[\protect\citeauthoryear{M{\"{u}}ller and Sch{\"{a}}fer}{2004}]{Muller2004}
    M{\"{u}}ller, H. H., Sch{\"{a}}fer, H. (2004). A General Statistical Principle for Changing a Design Any Time During the Course of a Trial. \textit{Statistics in Medicine} \textbf{23}, 2497--2508.
    
    \bibitem[\protect\citeauthoryear{R Core Team}{2015}]{r2015}
    {R Core Team} (2015). R: A Language and Environment for Statistical Computing. \textit{R Foundation for Statistical Computing}, Vienna, Austria.
    \url{http://www.R-project.org/}.
    
    \bibitem[\protect\citeauthoryear{Shuster}{2002}]{Shuster2002}
    Shuster, J. (2002). Optimal Two-Stage Designs for Single Arm Phase II Cancer Trials. \textit{Journal of Biopharmaceutical Statistics} \textbf{12}, 39--51.
    
    \bibitem[\protect\citeauthoryear{Simon}{1989}]{Simon1989}
    Simon, R. (1989). Optimal Two-Stage Designs for Phase II Clinical Trials. \textit{Controlled Clinical Trials} \textbf{10}, 1--10.
    
    \bibitem[\protect\citeauthoryear{Vazirani}{2001}]{Vazirani2001}
    Vazirani, V. V. (2001). Knapsack. In Approximation Algorithms pp. 68--73. Berlin: Springer.
    
    \bibitem[\protect\citeauthoryear{Wickham}{2009}]{ggplot2009}
    Wickham, H. (2009). ggplot2: Elegant Graphics for Data Analysis. \textit{Springer-Verlag} New York.
    
\end{thebibliography}

\end{document}